\documentclass[aps,prc,twocolumn,reprint,floatfix,superscriptaddress,showpacs]{revtex4-1}
\usepackage{graphicx}
\usepackage[pdftex,colorlinks,citecolor=blue,bookmarks]{hyperref}
\usepackage{bm}
\usepackage{longtable}
\begin{document}
% \draft
\title{Structure of Krypton isotopes calculated with symmetry conserving
configuration mixing methods}

\author{Tom\'as R. Rodr\'iguez} 
\affiliation{Departamento de F\'isica Te\'orica, Universidad Aut\'onoma de Madrid, E-28049 Madrid, Spain} 
\pacs{21.10.-k,21.60.Jz,27.50.+e}	
\begin{abstract}
Shape transitions and shape coexistence in the $^{70-98}$Kr region are studied in a unified view with state-of-the-art beyond self-consistent mean field methods based on the Gogny D1S interaction. Beyond mean field effects are taken into account through the exact angular momentum and particle number restoration and the possibility of axial and non-axial shape mixing. The results of the low-lying properties of these isotopes are in good agreement with the experimental data when the triaxial degree of freedom is included. Shape transitions from axial-oblate ($^{70-72}$Kr) to triaxial-prolate ($^{74-78}$Kr) and from spherical-triaxial ($^{86-92}$Kr) to axial-oblate ($^{94-98}$Kr) ground states are obtained. Additionally, low-lying $0^{+}$ excited states and quasi-gamma bands are found showing the richness of the collective structure in this region.
\end{abstract}
\maketitle
%%%%%%%%%%%%%%%%%%%%%%
%%%%%% INTRODUCTION %%%%%%%

%%%%%%%%%%%%%%%%%%%%%%
\section{Introduction}\label{intro}
%%%%%%%%%%%%%%%%%%%%%%
Shape evolution of Krypton isotopes have attracted a lot of experimental and theoretical interest in the past~\cite{RMP_83_1467_2011}. Such an evolution is rather complex and can be understood as a consequence of the various shell gaps found in the Nilsson single particle energies. The existence of these gaps produces energy landscapes with more than one equilibrium shape. Hence, the most energetically favored intrinsic configuration can change abruptly by adding neutrons, producing shape transitions along the isotopic chain. Furthermore, shape coexistence can occur if different configurations originate collective bands with $0^{+}$ band-heads close in energy. Additionally, these configurations can be mixed and the degree of mixing can be determined by the distortion of the rotational/vibrational behavior of the corresponding bands and by the transitions between states of different bands.\\ \indent  
In the neutron deficient side, fingerprints of shape coexistence as low-lying $0^{+}_{2}$ excited states~\cite{PRC_25_1941_1982,PRC_52_2444_1995,PRC_56_R2924_1997,PRL_90_082502_2003}, shape mixing as the distortion of low-lying bands~\cite{PRL_47_1514_1981}, as well as spectroscopic quadrupole moments and electromagnetic transitions~\cite{PRL_95_022502_2005,PRC_75_054313_2007,PRC_77_024312_2008} have been measured, supporting the possible coexistence of a prolate band and an oblate band for these nuclei. In addition, the experiments suggest that the shape transition occurs from an oblate $^{72}$Kr to prolate $^{74-76}$Kr ground states, although, based on recent Coulomb excitation measurements, a prolate character of the $2^{+}_{1}$ state in $^{72}$Kr have been proposed~\cite{Iwasaki_to_be_published}.\\ \indent
On the other hand, the onset of deformation in the neutron rich isotopes around the region $A\sim100$ is the subject of recent studies~\cite{PRL_108_062701_2012,NPA_899_1_2013}. Hence, while for Sr and Zr isotopes a quite sharp transition from spherical to deformed configurations in $N=58-60$ is observed, the neutron rich Kr isotopes do not show such a rapid change.  Furthermore, shape coexistence is found experimentally by the occurrence of two and three $0^{+}$ excited states below 2 MeV with strong $E0$ transitions in $^{96}$Sr and $^{98}$Zr, respectively ($N=58$)~\cite{NNDC,ADNDT_89_77_2005}. Unfortunately, there is no experimental data currently available for Kr isotopes in this region.\\ \indent
From the theoretical point of view, early self-consistent mean field studies have already shown a rich shape evolution in this region by analyzing the potential energy surfaces (PES) obtained with Skyrme~\cite{NPA_443_39_1985}, Relativistic~\cite{NPA_586_201_1995} and Gogny~\cite{PRL_62_2452_1989} interactions. However, spectroscopic information cannot be obtained within a pure mean field approach and several methods have been applied to calculate excitation energies, electromagnetic properties, etc. Due to the huge number of possible configurations, conventional large scale shell model calculations are still out of reach in this region. Nevertheless, results using the shell model Monte Carlo (SMMC) approach with a pairing plus quadrupole interaction~\cite{NPA_728_109_2003}, the VAMPIR method with an effective $G$-matrix defined in a reduced valence space~\cite{NPA_665_333_2000,NPA_770_107_2006} and the five dimensional collective hamiltonian (5DCH), also with a pairing plus quadrupole interaction~\cite{NPA_849_53_2011}, have been reported. \\ \indent
Additionally, there are several methods based on self-consistent underlying mean fields that have been applied to compute spectroscopic properties. Recently, a version of the Interacting Boson Model (IBM) has been used to study the shape dynamics of neutron rich Kr isotopes~\cite{NPA_899_1_2013}. In this case, the parameters of the IBM hamiltonian are found by mapping the mean field energy surfaces obtained with the Gogny D1M interaction to the corresponding IBM ones. The results reproduce quite well the available experimental data and they are consistent with a smooth triaxial-to-oblate transition around $N=60$. \\ \indent
On the other hand, the 5DCH method has been also used to study the low-lying spectroscopy in this region. Here, the inertial parameters and the potential energy of a Bohr collective hamiltonian are extracted from an underlying mean field, based on Gogny~\cite{PRC_75_054313_2007,PLB_676_39_2009} or Relativistic~\cite{PRC_87_054305_2013} interactions. Then, such a reduced problem is solved and the energy levels, moments and transitions are computed. The agreement with the available data is also good with both Gogny and Relativistic interactions. In particular, in the neutron deficient part, these calculations reveal the key role played by the triaxial degree of freedom to reproduce the correct deformation of the ground state and first excited state bands in $^{74-78}$Kr.\\ \indent
Finally, the most microscopic approaches consistent with the underlying mean field are the work of Bender \textit{et al.}~\cite{PRC_74_024312_2006} (with the Skyrme SLy6 interaction) and the calculation of $^{76}$Kr included in Refs.~\cite{PRC_87_054305_2013} (with the PC-PK1 relativistic lagrangian). There, the generator coordinate method (GCM) with particle number and angular momentum projected mean field states is applied. In both cases, contrary to the 5DCH results, these calculations do not reproduce the ordering of the low-lying levels, predicting oblate ground states also for $^{74-78}$Kr. Since the 5DCH is a gaussian overlap approximation of a GCM, a lack of triaxiality in these exact GCM calculations has been proposed as the most plausible explanation for this problem. Nevertheless, whether this contradiction is actually due to the inclusion of triaxial deformations or to the effective interaction, a study using the same underlying interaction and with the same beyond-mean-field approach, with and without including the triaxial degree of freedom, should be performed. This has been done recently in the nucleus $^{76}$Kr using the PC-PK1 relativistic lagrangian~\cite{PRC_89_054306_2014} but a more systematic study along the isotopic chain is still missing. \\ \indent
In this work the shape evolution from neutron deficient to stable and neutron rich Kr isotopes is studied in a unified manner, using the so-called Symmetry Conserving Configuration Mixing (SCCM) method. This framework is based on the GCM and includes quantum number restorations (particle number and angular momentum) and shape mixing of axial and triaxial intrinsic states. In addition, Gogny D1S~\cite{NPA_428_23_1984} is used as the underlying interaction at every step. Therefore, the triaxial degree of freedom is explored beyond mean field without using either IBM mappings or gaussian overlap approximations.\\ \indent
The paper is organized as follows. First, the most important aspects of the theoretical framework in Sec.~\ref{theo} are reviewed. Then, in Sec.~\ref{results}, the results of the calculations, starting from a mean field description in terms of the energy surfaces and Nilsson levels are shown. Next, the shape evolution along the isotopic chain and the role played by the triaxial degree of freedom are discussed with the help of the collective wave functions. Furthermore, both global (along the isotopic chain) and individual (nucleus by nucleus) theoretical results are compared with the experimental data. Finally, the most important conclusions of this work are summarized in Sec.~\ref{conclusions}.  
%%%%%%%%%%%%%%%%%%%%%%
\section{Theoretical framework}\label{theo}
%%%%%%%%%%%%%%%%%%%%%%
As stated above, in this paper the so-called Symmetry Conserving
Configuration Mixing (SCCM) method is used. A detailed description of this theoretical framework can be found in Refs.~\cite{PRC_81_064323_2010,PRC_78_024309_2008} (and references therein). Nevertheless, the most important aspects of this framework are now summarized, pointing out some differences with other similar methods used in the literature. 
For the sake of simplicity, the energy at different levels of complexity of the many-body method are written as expectation values of a hamiltonian in the following paragraphs. However, since the Gogny D1S interaction contains density dependencies, this notation is not fully rigorous and corresponding energy density functionals (EDF) should be defined otherwise. Furthermore, a proper definition of the EDF is also key to avoid the potential problems that come out when symmetry restorations and configuration mixing are performed within this framework (see Ref.~\cite{NPA_696_467_2001,PRC_79_044318_2009}  for a thorough description of this issue and Refs.~\cite{PRC_81_064323_2010,NPA_709_201_2002} for the definition of the present EDF).\\ \indent
In the SCCM method used in this work, the different many body states are calculated by mixing particle number and angular momentum restored intrinsic Hartree-Fock-Bogoliubov type wave functions (HFB) which have different quadrupole shapes (axial and non-axial)~\cite{PRC_81_064323_2010,PRC_78_024309_2008}:
\begin{equation}
|\Psi^{IM\sigma}\rangle=\sum_{\beta_{2},\gamma,K}f^{I\sigma}_{K}(\beta_{2},\gamma)P^{I}_{MK}P^{N}P^{Z}|\Phi(\beta_{2},\gamma)\rangle
\label{GCM_state}
\end{equation}
where $I$, $M$, $K$ are the total angular momentum and its projection on the $z$-axis in the laboratory and intrinsic frame respectively, $P^{I}_{MK}$ and $P^{N(Z)}$ the angular momentum and neutron (proton) projectors defined through integrals in the Euler and gauge angles respectively~\cite{RING_SCHUCK} and $\sigma$ labels different states obtained for a given value of $I$. The HFB-type states -$|\Phi(\beta_{2},\gamma)\rangle$- are found with the variation after particle number projection method (PN-VAP)~\cite{RING_SCHUCK,NPA_696_467_2001} i.e., the particle number projected energy is minimized imposing constraints in the quadrupole deformation ($\hat{Q}_{2\mu}=r^{2}Y_{2\mu}(\theta,\varphi)$):
\begin{eqnarray}
\delta\left(E'^{N,Z}(\beta_{2},\gamma)\right)&=&0\nonumber\\ \indent
E'^{N,Z}(\beta_{2},\gamma)=E^{N,Z}(\beta_{2},\gamma)&-&\lambda_{q_{20}}\langle\Phi|\hat{Q}_{20}|\Phi\rangle\nonumber\\ \indent&-&\lambda_{q_{22}}\langle\Phi|\hat{Q}_{22}|\Phi\rangle
\label{PN-VAP}
\end{eqnarray} 
where $\lambda_{q_{2\mu}}$ are Langrange multipliers that guarantee the conditions:
\begin{eqnarray}
\lambda_{q_{20}}\rightarrow\langle\Phi|\hat{Q}_{20}|\Phi\rangle=q_{20}\nonumber\\ \indent
\lambda_{q_{22}}\rightarrow\langle\Phi|\hat{Q}_{22}|\Phi\rangle=q_{22} 
\end{eqnarray}
In addition, the deformation parameters $(\beta_{2},\gamma)$ are directly related to $(q_{20},q_{22})$ by:
\begin{eqnarray}
q_{20}=\frac{\beta_{2}\cos\gamma}{C}\,\,;\,\,q_{22}=\frac{\beta_{2}\sin\gamma}{\sqrt{2}C}\,\,;\,\,C=\sqrt{\frac{5}{4\pi}}\frac{4\pi}{3r_{0}^{2}A^{5/3}}
\label{betagamma}
\end{eqnarray}
being $r_{0}=1.2$ fm and $A$ the mass number.
The PN-VAP energy in Eq.~\ref{PN-VAP} defines a potential energy surface (PES) in the $(\beta_{2},\gamma)$ plane which is useful to analyze the intrinsic shape of the nucleus:
\begin{equation}
E^{N,Z}(\beta_{2},\gamma)=\frac{\langle\Phi(\beta_{2},\gamma)|\hat{H}P^{N}P^{Z}|\Phi(\beta_{2},\gamma)\rangle}{\langle\Phi(\beta_{2},\gamma)|P^{N}P^{Z}|\Phi(\beta_{2},\gamma)\rangle}
\label{PN-VAP-PES}
\end{equation}
The states in Eq.~\ref{GCM_state} are the generator coordinate method (GCM) ansatz. Hence, the parameters $f^{I\sigma}_{K}(\beta_{2},\gamma)$ are variational parameters that are found by solving the Hill-Wheeler-Griffin (HWG) equations~\cite{RING_SCHUCK}:
\begin{equation}
\sum_{\beta'_{2}\gamma'K'}
\left(\mathcal{H}^{I}_{\beta_{2}\gamma K,\beta'_{2}\gamma'K'}-E^{I\sigma}\mathcal{N}^{I}_{\beta_{2}\gamma K,\beta'_{2}\gamma'K'}\right)f^{I\sigma}_{K'}(\beta'_{2},\gamma')=0
\label{RitzGCM}
\end{equation}
The energy and norm overlap matrices are defined as:
\begin{eqnarray}
\mathcal{H}^{I}_{\beta_{2}\gamma K,\beta'_{2}\gamma'K'}&=&\langle\Phi(\beta_{2},\gamma)|\hat{H}P^{I}_{KK'}P^{N}P^{Z}|\Phi(\beta'_{2},\gamma')\rangle\nonumber\\ \indent
\mathcal{N}^{I}_{\beta_{2}\gamma K,\beta'_{2}\gamma'K'}&=&\langle\Phi(\beta_{2},\gamma)|P^{I}_{KK'}P^{N}P^{Z}|\Phi(\beta'_{2},\gamma')\rangle
\end{eqnarray}
Normally the HWG equation (one for each value of $I$) is solved by transforming the generalized eigenvalue problem defined by Eq.~\ref{RitzGCM} into a regular diagonalization problem~\cite{RING_SCHUCK}. To do so, the norm overlap matrix is first diagonalized:
\begin{equation}
\sum_{\beta'_{2}\gamma'K'}\mathcal{N}^{I}_{\beta_{2}\gamma K,\beta'_{2}\gamma'K}U^{I}_{\Lambda;\beta'_{2}\gamma'K'}=n^{I}_{\Lambda}U^{I}_{\Lambda;\beta_{2}\gamma K}
\end{equation}
Then, an orthonormal set of states, the so-called natural basis, is built with the eigenvectors and eigenvalues of the norm overlap:
\begin{equation}
|\Lambda^{IM}\rangle=\sum_{\beta_{2}\gamma K}\frac{U^{I}_{\Lambda;\beta_{2}\gamma K}}{\sqrt{n^{I}_{\Lambda}}}P^{I}_{MK}P^{N}P^{Z}|\Phi(\beta_{2},\gamma)\rangle\,\,;n^{I}_{\Lambda}\neq0
\end{equation}
In the last equation, the linear dependencies of the original set of states are removed by choosing those norm eigenvalues that are different from zero. Therefore, both the GCM ansatz (Eq.~\ref{GCM_state}) and HWG equations (Eq.~\ref{RitzGCM}) can be rewritten as:
\begin{eqnarray}
|\Psi^{IM\sigma}\rangle&=&\sum_{\Lambda}g^{I\sigma}_{\Lambda}|\Lambda^{IM}\rangle\\ \indent
\sum_{\Lambda'}\langle\Lambda^{IM}|\hat{H}|\Lambda'^{IM}\rangle g^{I\sigma}_{\Lambda'}&=&E^{I\sigma}g^{I\sigma}_{\Lambda}
\end{eqnarray}
The solution of the HWG equations provide the coefficients $g^{I\sigma}_{\Lambda}$, from which observables such as energy spectrum, radii, electromagnetic moments, reduced transition probabilities, etc. can be calculated. In addition, the weights of a given intrinsic $(\beta_{2},\gamma)$ configuration in the corresponding GCM state, the so-called collective wave functions, are also given as a function of the coefficients defined above:
\begin{equation}
G^{I\sigma}(\beta_{2},\gamma)=\sum_{K,\Lambda}g^{I\sigma}_{\Lambda}U^{I}_{\Lambda;\beta_{2}\gamma K}
\label{coll_wf}
\end{equation}
These quantities are useful to understand the intrinsic structure of the ground and excited states in terms of these collective coordinates.\\ \indent
All of the above expressions can be largely simplified if the HFB-type states are axially symmetric~\cite{PRL_99_062501_2007}. In particular, $\gamma$ only takes two values, namely, $\gamma=0^{\circ}$ (prolate, $\beta_{2}>0$) and $\gamma=60^{\circ}$ (oblate, or, equivalently, $180^{\circ}$, $\beta_{2}<0$).  The angular momentum projection is then reduced to $K=0$ values and only one of the three integrals in the Euler angles has to be evaluated. 
To check the relevance of the triaxial degree of freedom in the Krypton isotopic chain, both axial and triaxial calculations with the same underlying interaction -Gogny D1S- are discussed throughout this document.\\ \indent
%%%%%%%%%%%%%%%%%%%%%%%%
\begin{figure*}[t]
\begin{center}
  \includegraphics[width=1.0\textwidth]{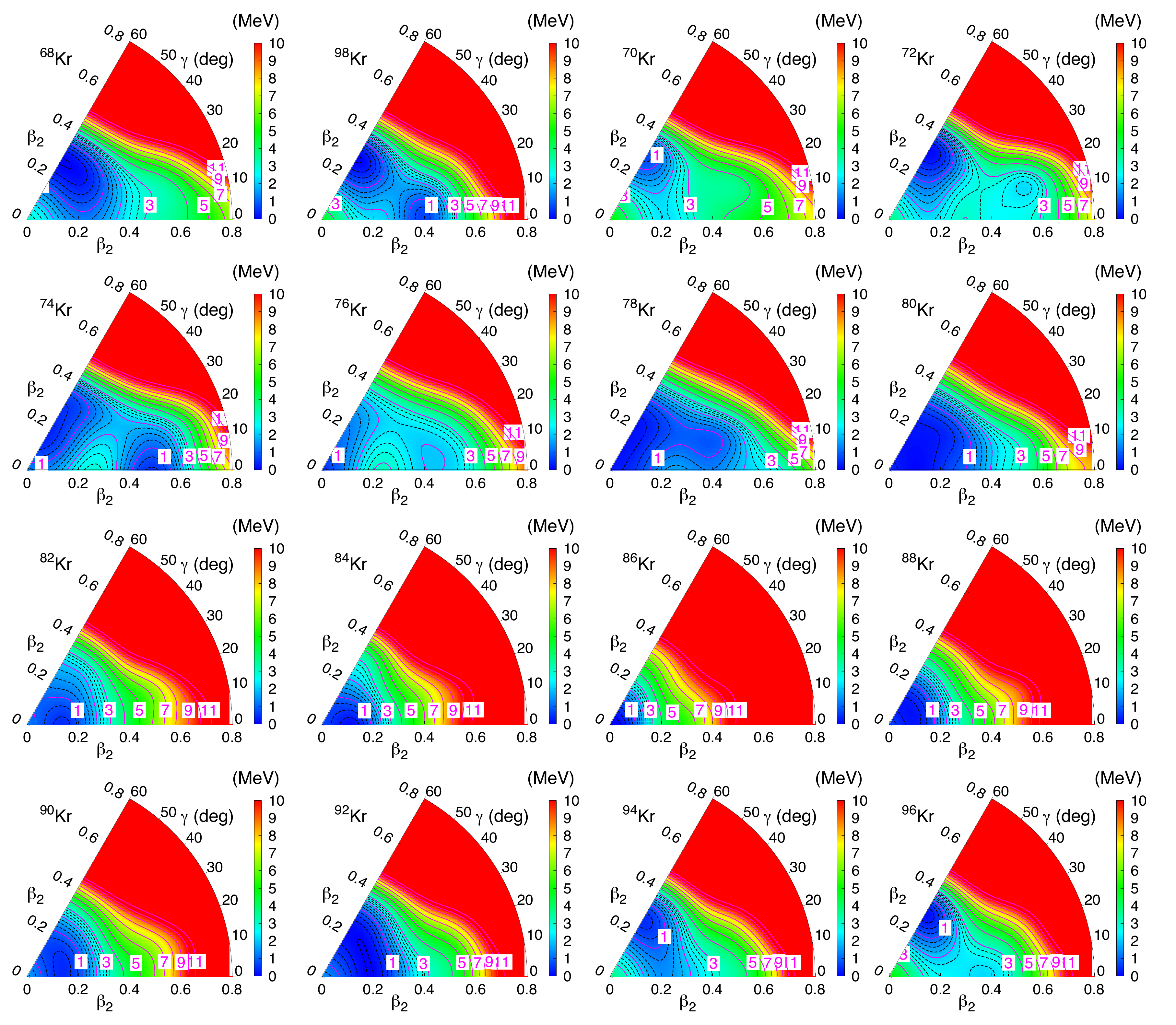}
\end{center}
\caption{(color online) Particle number projected potential energy surface in the $(\beta,\gamma)$ plane for $^{68-98}$Kr. Contour lines are separated 1 MeV (solid lines) and 0.25 MeV (dashed lines) and the energies are normalized to the minimum of each surface.}\label{PES_VAP}
\end{figure*}
%%%%%%%%%%%%%%%%%%%%%%%%
One should mention the methodological main differences between the present study and some recent calculations reported in the Kr isotopic chain. As stated in the introduction, generator coordinate method with exact particle number and angular momentum projection calculations have been already performed with Skyrme~\cite{PRC_74_024312_2006} and Relativistic~\cite{PRC_87_054305_2013} energy density functionals but including only axial shapes $(\gamma=0^{\circ},180^{\circ})$ which is a strong limitation to study shape transitions and coexistence in this region. Only the nucleus $^{76}$Kr has been recently computed considering the triaxial degree of freedom with the relativistic framework~\cite{PRC_89_054306_2014}. Nevertheless, in those cases, plain HFB or Lipkin-Nogami (LN) intrinsic wave functions are used instead of the ones provided by the PN-VAP method. Hence, pairing correlations are much better described than in the HFB and LN approaches that cannot account for such correlations in weak pairing regimes~\cite{PLB_545_62_2002,PRL_99_062501_2007}. On the other hand, five dimensional collective hamiltonian (5DCH) results have been reported using the Gogny D1S~\cite{PRC_75_054313_2007,PLB_676_39_2009} and Covariant~\cite{PRC_87_054305_2013} interactions which include quadrupole triaxial shapes. However, the 5DCH is deduced from the GCM method assuming a gaussian overlap approximation (GOA, see Ref.~\cite{EPJA_9_35_2000}). Quantum number projections are not taken into account in such a method either. This fact could affect significantly the spectrum predicted by 5DCH calculations since the absence of particle number restoration leads to spurious mixing of solutions with different number of particles~\cite{PLB_704_520_2011,PRC_88_064311_2013}. 
Finally, recent calculations have been reported in this region within the interacting boson model (IBM) framework~\cite{PRL_108_062701_2012,NPA_899_1_2013}. Here the IBM hamiltonian of each isotope is mapped to the HFB potential energy surface in the $(\beta_{2},\gamma)$ plane calculated with the Gogny D1M interaction~\cite{PRL_102_242501_2009}. Then, theoretical predictions are obtained after solving the IBM hamiltonian in a restricted valence space. \\ \indent
Consequently, in contrast to the methods reported above, the method used here includes the triaxial degree of freedom, quantum restoration and shape mixing correlations self-consistently and they are free from GOA approaches. Nevertheless, there are some limitations -shared also with the calculations already mentioned- that must be pointed out. \\ \indent
The most important limitations concern to HFB-type states, which do not contain explicit quasiparticle excitations, conserve both time-reversal and spatial parity symmetries and do not allow the inclusion of proton-neutron pairing. Thus, neither negative parity states (the lowest experimental $3^{-}$ state is found in $^{72}$Kr at 1.85 MeV~\cite{NNDC}) nor pure single particle excitations can be obtained. Nevertheless, parity-breaking calculations in this region with Gogny interactions have shown in all of the nuclei studied here a static octupole deformation equal to zero~\cite{PRC_84_054302_2011}. Therefore, its influence on the spectra should be eventually considered through octupole fluctuations and parity projection which are beyond the scope of the present work.\\ \indent
Furthermore, excited states are not explored in a fully efficient way from the variational point of view due to the above restrictions. The inclusion of other degrees of freedom such as pairing~\cite{PLB_704_520_2011,PRC_88_064311_2013} or quadrupole fluctuations~\cite{PRC_71_044313_2005} would improve the variational description of the excited states, particularly the $0^{+}$ states, but they are not considered here since they would increase prohibitively the computational time. Finally, since the HFB-type states have a product structure of protons and neutrons separately, $T_{z}=0$ pairing is not taken into account in the present calculations. This limitation can be particularly relevant in describing the $N\approx Z$ nuclei but, again, an isospin mixing~\cite{PRL_87_052504_2001} performed in combination with the present SCCM method is beyond the present study. Nevertheless, the degrees of freedom relevant to describe qualitatively and, to some extent, also quantitatively the low-lying states in the Kr isotopic chain are expected to be included in the present work.\\ \indent
Concerning some technical details about the calculations, a regular triangular mesh in the triaxial plane including $N_{\mathrm{GCM}}=60$ HFB-type states are used. Each of these states is expanded in a single particle basis with nine major spherical harmonic oscillator shells. The number of integration points in the Euler -$(a,b,c)$- and gauge -$\varphi$- angles are chosen to ensure the convergence of both diagonal and non-diagonal projected matrix elements of the total angular momentum and particle number operators to the nominal values $I(I+1)$, $N$ and $Z=36$ respectively. In this case, these values are $N_{a}=8$, $N_{b}=N_{c}=16$ and $N_{\varphi}=9$ (see Ref.~\cite{PRC_81_064323_2010} for details). The calculations were performed at the high performance computing facility Prometheus at GSI (Darmstadt)~\cite{GSI} with a computing time of $~48000 h$ in a single processor for each nucleus. 
%%%%%%%%%%%%%%%%%%%%%%
\section{Results and discussion}\label{results}
%%%%%%%%%%%%%%%%%%%%%%
\subsection{PN-VAP potential energy surfaces and Nilsson-like single particle energies}
%%%%%%%%%%%%%%%%%%%%%%
%%%%%%%%%%%%%%%%%%%%%%%%
\begin{figure}[t]
\begin{center}
  \includegraphics[width=0.8\columnwidth]{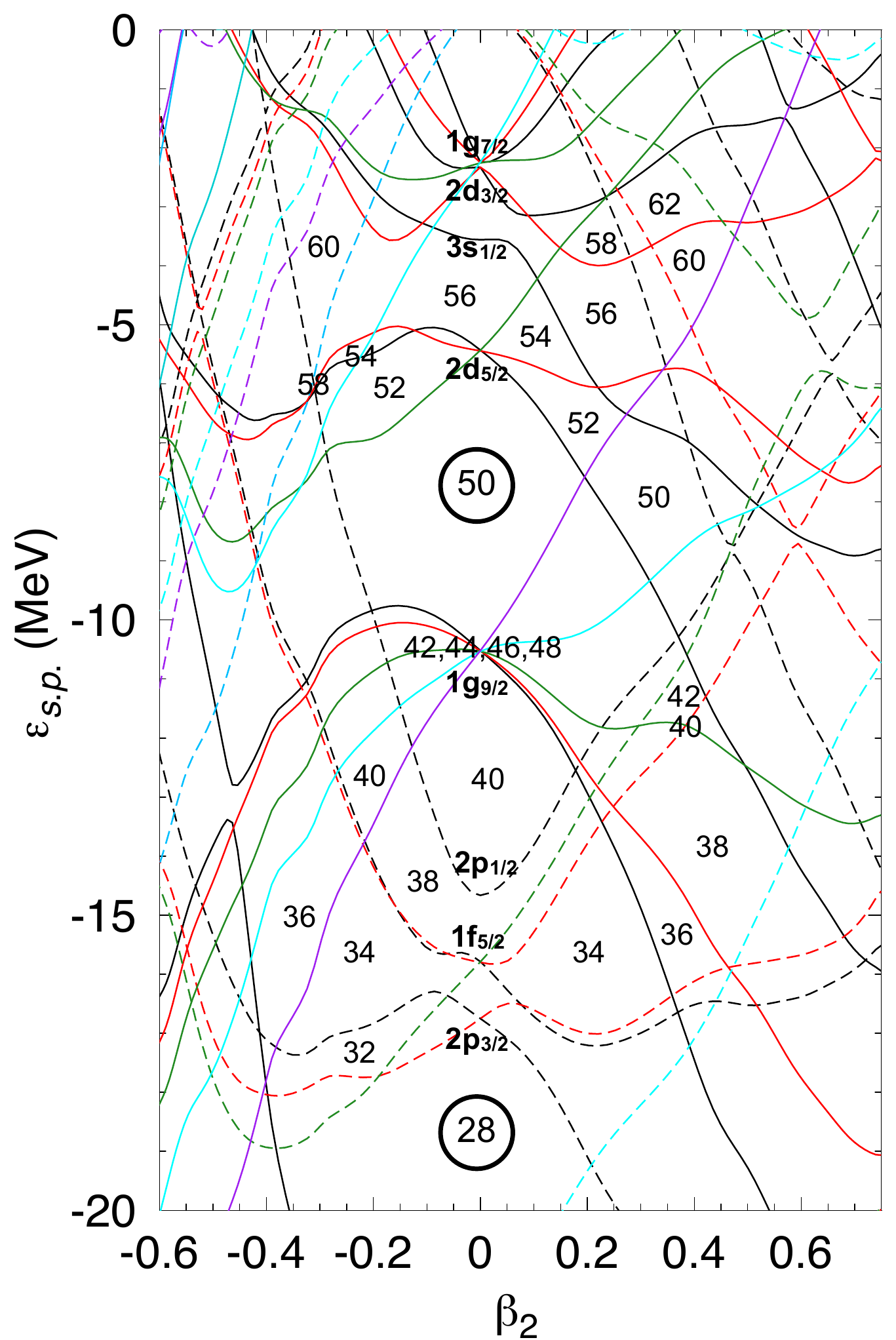}
\end{center}
\caption{(color online) Neutron single particle energies as a function of the axial quadrupole deformation parameter calculated for $^{96}$Kr with Gogny D1S interaction.  Neutron number in the gaps and level crossings indicate the minima in the axial potential energy surfaces found for each nuclei ($^{68-98}$Kr). Continuous (dashed) lines correspond to positive (negative) levels and the color code represent the value of $j_{z}$: 1/2, 3/2, 5/2, 7/2, 9/2, 11/2 match with black, red, green, cyan, purple and light blue respectively.}\label{Nilsson}
\end{figure}
%%%%%%%%%%%%%%%%%%%%%%%%
%%%%%%%%%%%%%%%%%%%%%%%%
\begin{figure*}[t]
\begin{center}
  \includegraphics[width=\textwidth]{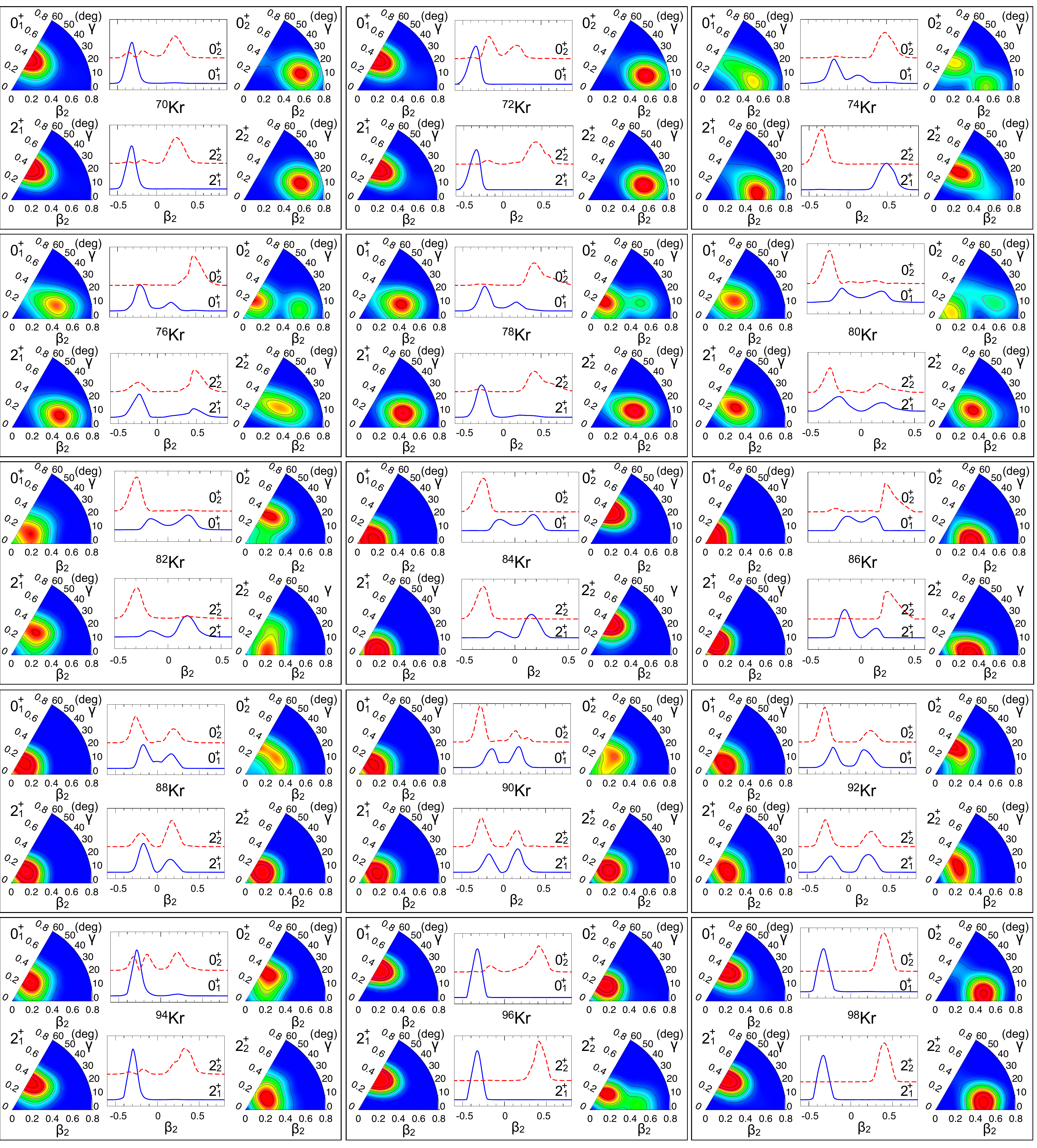}
\end{center}
\caption{(color online) Collective wave functions for the ground state ($0^{+}_{1}$) and $2^{+}_{1}$, $0^{+}_{2}$ and $2^{+}_{2}$ excited states calculated with the SCCM method for $^{70-98}$Kr isotopes (from left to right and from top to bottom). One-dimensional plots represent axial calculations while 'pie-like' plots represent full triaxial results (color scale: red and blue mean large and small height respectively).}
\label{WF_ALL}
\end{figure*}
%%%%%%%%%%%%%%%%%%%%%%%%
%%%%%%%%%%%%%%%%%%%%%%%%
\begin{figure}[t]
\begin{center}
  \includegraphics[width=0.75\columnwidth]{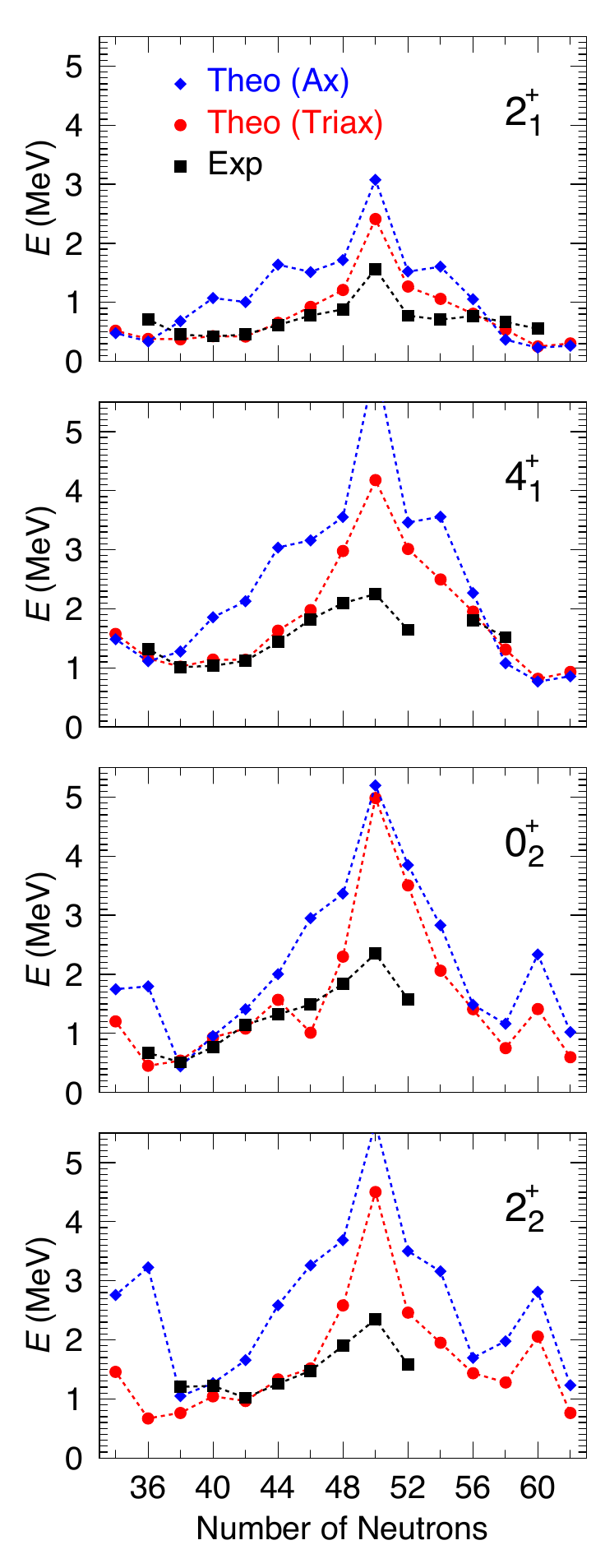}
\end{center}
\caption{(color online) Excitation energies along the Krypton isotopic chain $^{70-98}$Kr for (a) $2^{+}_{1}$, (b) $4^{+}_{1}$, (c) $0^{+}_{2}$, and (d) $2^{+}_{2}$ states. Black boxes, blue diamonds and red bullets represent the experimental values (taken from~\cite{NNDC}), and the results of SCCM axial and SCCM triaxial calculations respectively.}
\label{exc_ener}
\end{figure}
A first physical insight on the shape of the Kr isotopes can be obtained from the potential energy surfaces (PES, see Eq.~\ref{PN-VAP-PES}) plotted in Fig.~\ref{PES_VAP}. This figure reveals the large variety of energy landscapes found in this isotopic chain with examples of spherical, oblate, prolate, $\gamma$-soft and oblate-prolate shape-coexistent nuclei. 
Starting from the lighter isotopes, single absolute oblate minima are obtained for $^{68-70}$Kr, having for the former a noticeable $\gamma$-softness. For $^{72-76}$Kr two clear minima are observed in the PES; oblate and triaxial/prolate in $^{72}$Kr; axial oblate and prolate in $^{74}$Kr, being both almost degenerated; and spherical and prolate in $^{76}$Kr.  The latter two minima tend to merge in $^{78}$Kr and for $^{80}$Kr a large degeneracy around the spherical shape up to $\beta_{2}=0.3$ is observed. Such a degeneracy is drastically reduced in $^{82}$Kr where a slightly prolate single minimum is found, similarly to $^{84}$Kr. After the spherical semi-magic nucleus $^{86}$Kr and slightly spherical $^{88-90}$Kr isotopes, a $\gamma$-soft nucleus, $^{92}$Kr, is obtained. Then, for $^{94}$Kr a single oblate/$\gamma$-soft minimum is observed and, finally, potential energy surfaces with two axial minima (oblate and prolate) are found for the neutron rich isotopes $^{96-98}$Kr. \\ \indent
The general behavior of the energy landscapes and the position of the minima can be understood by analyzing the underlying single particle levels. Hence, these minima correspond to the appearance of shell gaps and/or level crossings in a Nilsson-like spectrum. In Fig.~\ref{Nilsson} the single particle energies calculated self-consistently for neutrons in $^{96}$Kr are represented. Similar spectra -although shifted in energy- are obtained both for protons and for the rest of the isotopes in the chain. For the sake of simplicity, the axial quadrupole direction is plotted only. \\ \indent 
First, the structure of the protons for the Kr isotopes ($Z=36$) is described. Hence, the proton Fermi energy crosses two gaps, oblate ($\beta_{2}\sim-0.30$) and prolate ($\beta_{2}\sim+0.35$). Both gaps are produced by the filling -on top of the magic number $Z=28$- of the two $2p_{3/2}$ sub-shells, another level coming from $1f_{5/2}$ and one from $1g_{9/2}$ sub-shells. Therefore, the configuration of protons in the Krypton isotopes favor the appearance of oblate and prolate deformation.\\ \indent
The role of the neutrons is now analyzed. One can see in Fig.~\ref{Nilsson} that the structure of the neutron deficient and stable Krypton isotopes ($^{68-86}$Kr) is dominated by the $pf$ and $1g_{9/2}$ shells. Hence, the oblate shapes found at $^{68-70}$Kr can be associated both to the structure of the protons discussed above and the appearance of gaps between the $2p_{3/2}$ and $1f_{5/2}$ levels; the shape coexistence in $^{72-76}$Kr is related both to the spherical gaps between $1f_{5/2}-2p_{1/2}$ ($N=38$) and $2p_{1/2}-1g_{9/2}$ ($N=40$) shells, and the filling of the $1g_{9/2}$ and emptying of the $pf$ sub-shells. The nuclei $^{78-86}$Kr are dominated by the filling of the spherical $1g_{9/2}$ shell.\\ \indent
Above $N=50$, several cases are identified, i.e., nuclei: close to sphericity/$\gamma$-soft, $^{88-92}$Kr, associated to gaps produced by the $2d_{5/2}$ and $3s_{1/2}$ shells; with oblate minima, $^{94-98}$Kr, due to the proton gap and the filling of $1h_{11/2}$ levels with the highest values of $j_{z}$; and with prolate minima, coexisting with the oblate ones, $^{96-98}$Kr, given by the lowest $j_{z}$ levels from $1h_{11/2}$ that are crossing below the neutron Fermi level.
%%%%%%%%%%%%%%%%%%%%%%
\subsection{Collective wave functions}
%%%%%%%%%%%%%%%%%%%%%%%%
%%%%%%%%%%%%%%%%%%%%%%%%
\begin{figure}[t]
\begin{center}
  \includegraphics[width=0.75\columnwidth]{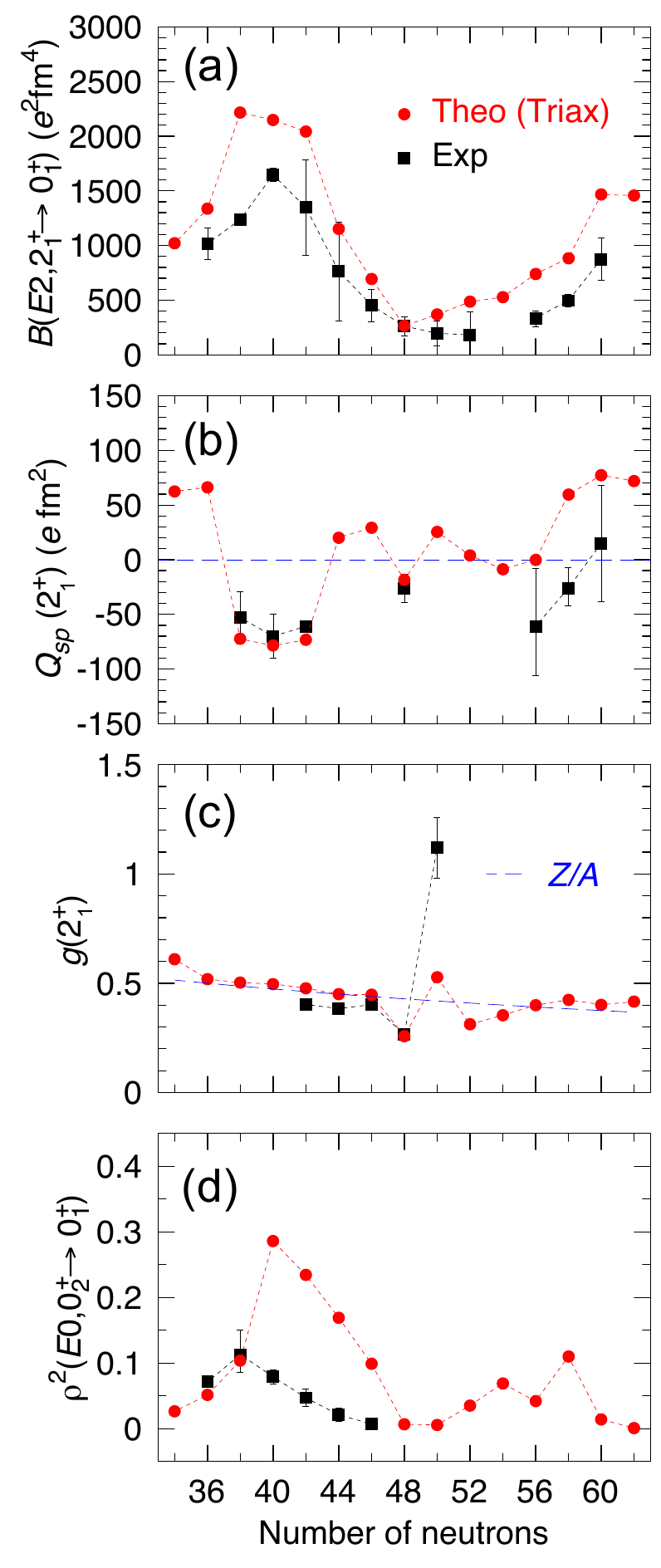}
\end{center}
\caption{(color online) (a) Electric quadrupole $(E2)$ reduced transition probabilities between $2^{+}_{1}$ and $0^{+}_{1}$ states. 
(b) Spectroscopic electric quadrupole moments and (c) Gyromagnetic factors for $2^{+}_{1}$ states. (d) Electric monopole $E0$ transition strength between $0^{+}_{2}$ and $0^{+}_{1}$ states. Black boxes and red bullets represent the experimental values (taken from~\cite{PRC_75_054313_2007,NPA_899_1_2013,NPA_700_59_2002,NPA_770_107_2006,PRC_64_024314_2001,NNDC,PRL_90_082502_2003,PRC_47_521_1993,PRC_31_1483_1985,ADNDT_89_77_2005,PLB_546_48_2002}) and the results of SCCM triaxial calculations respectively. Blue dashed line in (c) is the $Z/A$ line provided by the collective rotor model~\cite{RING_SCHUCK}.}
\label{BE2_Q21_g21_RHO0}
\end{figure}
%%%%%%%%%%%%%%%%%%%%%%%%
%%%%%%%%%%%%%%%%%%%%%%%%
\begin{figure*}[t]
\begin{center}
  \includegraphics[width=\textwidth]{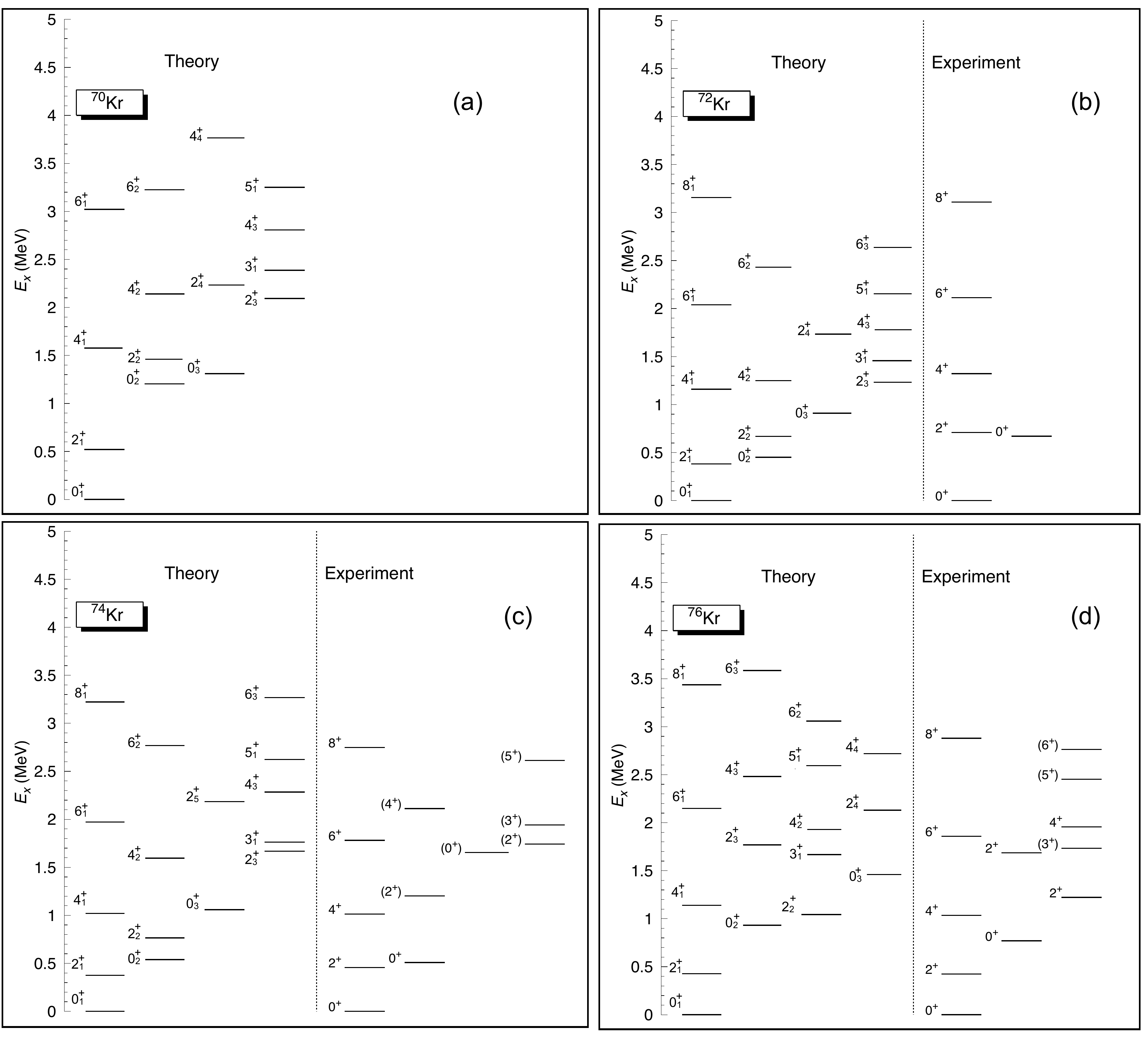}
\end{center}
\caption{Excitation energies for $^{70-76}$Kr isotopes. Experimental values are taken from Ref.~\cite{NNDC}.}
\label{Fig70_76}
\end{figure*}
%%%%%%%%%%%%%%%%%%%%%%%%
%%%%%%%%%%%%%%%%%%%%%%%%
\begin{figure*}[t]
\begin{center}
  \includegraphics[width=\textwidth]{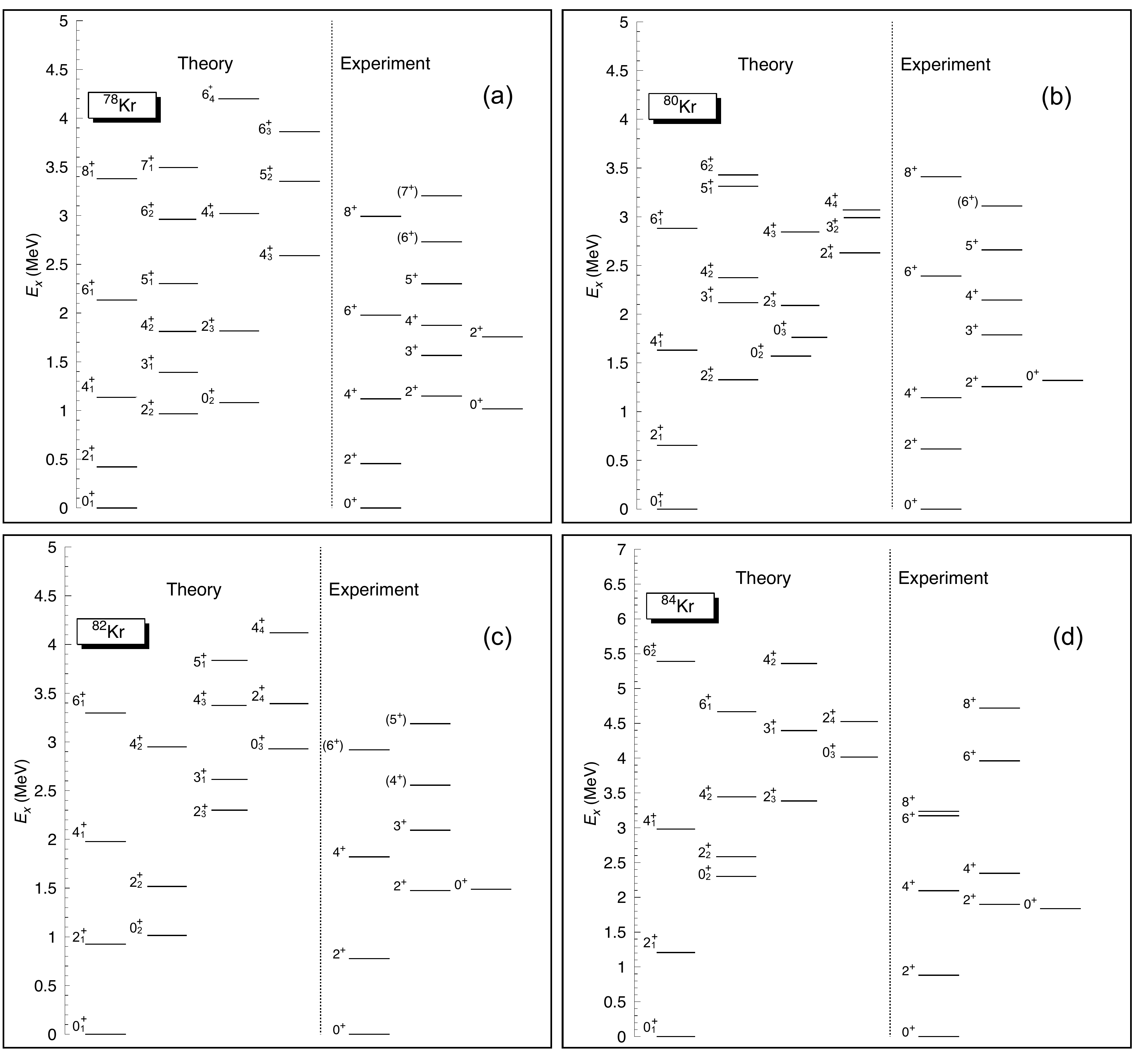}
\end{center}
\caption{Same as Fig.~\ref{Fig70_76} but for $^{78-84}$Kr isotopes.}
\label{Fig78_84}
\end{figure*}
%%%%%%%%%%%%%%%%%%%%%%%%  
%%%%%%%%%%%%%%%%%%%%%%%%
\begin{figure*}[t]
\begin{center}
  \includegraphics[width=\textwidth]{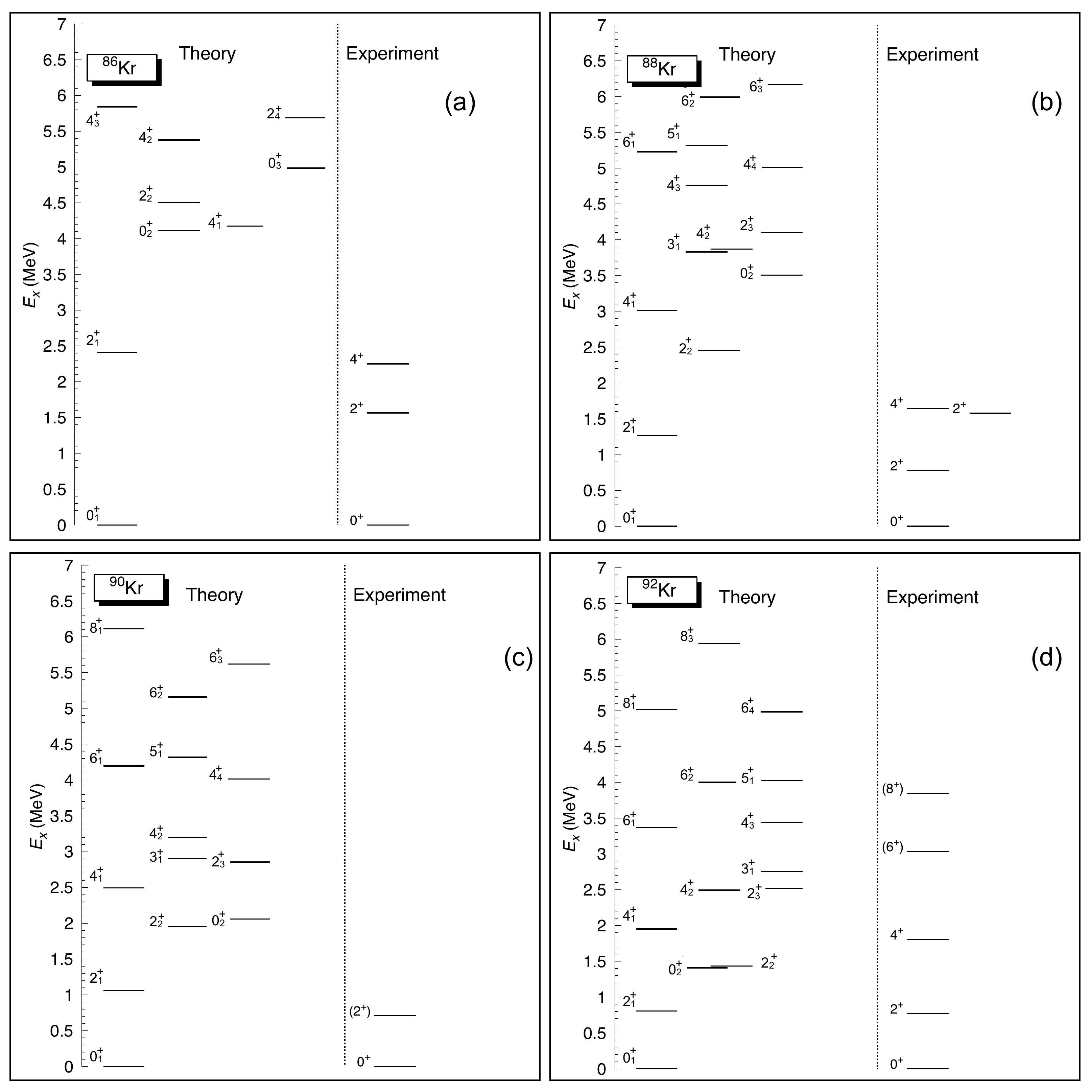}
\end{center}
\caption{Same as Fig.~\ref{Fig70_76} but for $^{86-92}$Kr isotopes.}
\label{Fig86_92}
\end{figure*}
%%%%%%%%%%%%%%%%%%%%%%%%  
%%%%%%%%%%%%%%%%%%%%%%%%
\begin{figure*}[t]
\begin{center}
  \includegraphics[width=\textwidth]{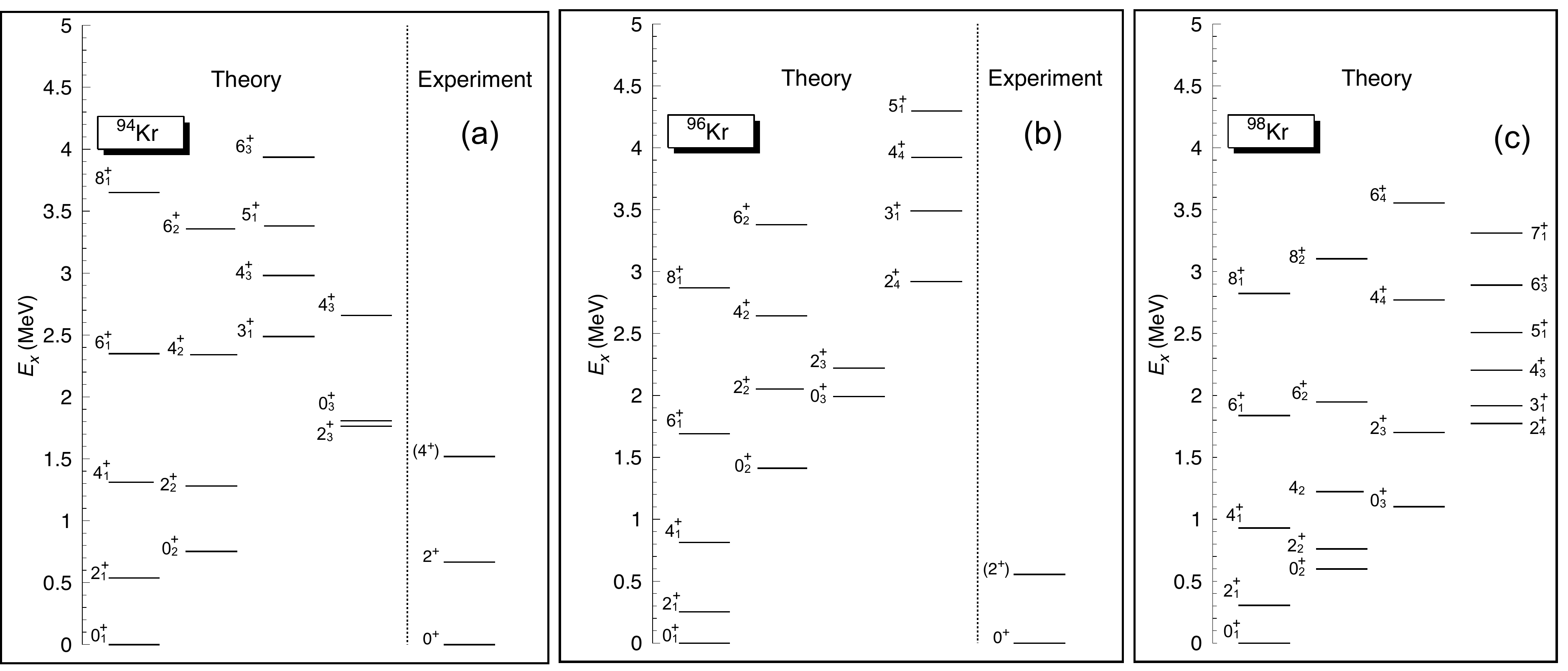}
\end{center}
\caption{Same as Fig.~\ref{Fig70_76} but for $^{94-98}$Kr isotopes.}
\label{Fig94_98}
\end{figure*}
%%%%%%%%%%%%%%%%%%%%%%%%  
The analysis given in the preceding section can be considered as a 'mean-field' based exploration of the structure of the nuclei studied in this work. However, the shape of every single nucleus can be examined state by state after performing the symmetry restorations and shape mixing within the SCCM method sketched in Sec.~\ref{theo}. Hence, the so-called collective wave functions (c.w.f. from now on) represent the weights of each intrinsic deformation in building the many-body nuclear states $|\Psi^{IM\sigma}\rangle$~\cite{RING_SCHUCK}. Since one of the aims of the present work is the analysis of the role played by the triaxial degree of freedom, both axial and triaxial calculations are performed. Obviously, the former are very much lighter than the latter in terms of computing time. In Fig.~\ref{WF_ALL} the axial and triaxial c.w.f. for the first two $0^{+}$ and $2^{+}$ states in $^{70-98}$Kr are shown. In each box, that corresponds to a given nucleus, $0^{+}_{1}$ and $0^{+}_{2}$ are shown in the upper part  and $2^{+}_{1}$ and $2^{+}_{2}$ in the lower part. Additionally, the axial results are plotted in the middle panel and the triaxial ones in the left (yrast states) and right (lowest $0^{+}_{2}$ and $2^{+}_{2}$ states) panels.\\ \indent
Before entering into the details of the shape evolution along the isotopic chain, a general remark about the differences/similarities between the axial and triaxial calculations should be made. Comparing both approaches, the axial c.w.f. can be interpreted in many cases just as the reduction to one degree of freedom of the triaxial ones. This is the case for all the states represented in Fig.~\ref{WF_ALL} for $^{84-92,98}$Kr, the $0^{+}_{1}$, $2^{+}_{1}$ states for $^{70,72,94,96}$Kr, the $2^{+}_{1}$, $2^{+}_{2}$ states for $^{74}$Kr, and the $0^{+}_{1}$, $0^{+}_{2}$ states for $^{82}$Kr. For the rest of states, the axial and triaxial c.w.f. are either of different nature and/or the ordering is exchanged. This is particularly important in the neutron deficient region -$^{74-80}$Kr- where pure axial SCCM calculations had problems in reproducing the experimental data in contrast to 5DCH calculations including triaxiality~\cite{PRC_74_024312_2006,PLB_676_39_2009,PRC_87_054305_2013}. This issue is discussed in detail below. 
One should mention a final remark about the axial/triaxial comparison. Although some of the axial collective functions can be associated with their triaxial partners, the quality of the axial approach could be different depending on the nucleus. Hence, more similar results are obtained in cases where the triaxial degree of freedom does not play a role -$^{98}$Kr, for example- than in those where pure triaxial states are found and the axial c.w.f. present a symmetric double peak structure around the spherical point -$^{90-92}$Kr, for example.\\ \indent
The shape evolution along the isotopic chain is discussed now. Starting from the neutron deficient part, axial oblate $(\beta_{2}\sim-0.35)$ $0^{+}_{1}$ and $2^{+}_{1}$ states for $^{70-72}$Kr are obtained. The triaxial results for $0^{+}_{2}$ and $2^{+}_{2}$ are also similar in both nuclei with the triaxial/prolate c.w.f. peaked in $(\beta_{2},\gamma)\sim(0.6,15^{\circ})$. In contrast, the axial results present a slightly less deformed prolate states for $^{70}$Kr and an almost symmetric prolate/oblate shape mixing ($0^{+}_{2}$) and a prolate ($2^{+}_{2}$) excited states for $^{72}$Kr.\\ \indent
As stated above, the most significant differences between axial and triaxial calculations are found in the structure of the $^{74-80}$Kr isotopes. For $^{74}$Kr a ground state c.w.f. that is quite extended in the $\gamma$ direction is obtained. Its maximum corresponds to a triaxial/prolate deformation -$(0.5,10^{\circ})$. The $2^{+}_{1}$ state peaks also in a similar deformation although the c.w.f. is more condensed around its maximum. On the other hand, the first $0^{+}$ excited state shows a clear shape mixing of oblate -$(0.35,60^{\circ})$- and prolate -$(0.5,0^{\circ})$- shapes, having the former a larger contribution. Finally, the $2^{+}_{2}$ c.w.f. peaks in the oblate part of the $(\beta_{2},\gamma)$ plane also at -$\sim(0.35,60^{\circ})$. Therefore, the structure of the $0^{+}$ states corresponds to the mixing through the $\gamma$ direction of the two minima observed in the PES (see  Fig.~\ref{PES_VAP}) while the $2^{+}$ states are less mixed and more constrained inside the potential wells. Looking at the axial calculation for this nucleus, basically the same result as the triaxial one for the $2^{+}$ states is found but the $0^{+}$ states contain a smaller mixing and are inverted with respect to the triaxial calculation. Consequently, the impossibility of shape mixing through the triaxial degree of freedom in axial calculations affects significantly the structure obtained within this approximation. \\ \indent
Similar conclusions can be extracted for $^{76-78}$Kr isotopes, where the $\gamma$-softness that connects different oblate and oblate shapes allows the mixing in the $\gamma$ direction. The ground and $2^{+}_{1}$ states are peaked in triaxial/prolate configurations; the $0^{+}_{2}$ c.w.f. present two maxima, the absolute one in an axial oblate shape and the other maximum in a prolate ($^{76}$Kr) state or a triaxial/prolate ($^{78}$Kr) state; and the $2^{+}_{2}$ are quite extended in the triaxial plane and could be considered as the band-heads of quasi-$\gamma$ bands. The axial calculations are rather different since oblate ground state bands and prolate excited state bands are obtained for both nuclei. 
These results for the neutron deficient isotopes are fully consistent with the axial SCCM calculations performed with the Skyrme SLy6 interaction~\cite{PRC_74_024312_2006} and with the triaxial 5DCH calculations carried out both with the Gogny D1S~\cite{PRC_75_054313_2007,PLB_676_39_2009} and Relativistic PC-PK1~\cite{PRC_87_054305_2013} interactions. However, in the present case, the same effective interaction and the same many-body method is used both for axial and triaxial calculations and the only difference between them is whether the triaxial degree of freedom is included. From the present calculations one can safely conclude that the disagreement with the experimental data obtained with an axial SCCM method~\cite{PRC_74_024312_2006} is more related to the lack of the triaxial degree of freedom rather than a drawback of the underlying effective interaction. This result is also obtained for the nucleus $^{76}$Kr calculated with a SCCM method based on a Relativistic functional~\cite{PRC_89_054306_2014}.\\ \indent
Moving towards the $N=50$ shell closure, a smooth transition in the ground state c.w.f. with a dominant role of the triaxial degree of freedom is observed. Hence, the shape evolution from the triaxial/prolate $^{78}$Kr to the spherical  $^{86}$Kr ground states proceeds through triaxial/oblate ($^{80}$Kr), pure triaxial ($^{82}$Kr) and prolate/spherical ($^{84}$Kr) configurations. On the other hand, the evolution of the excited states is not smooth. The calculations show $2^{+}_{1}$ states with dominant triaxial/oblate shapes for $^{80-82}$Kr, axial prolate shapes for $^{84}$Kr and axial oblate shapes for $^{86}$Kr. However, the main difference appears in the shapes of the $0^{+}_{2}$ and $2^{+}_{2}$ in this interval of nuclei. For $^{84}$Kr and $^{86}$Kr, an axial oblate and an axial prolate rotational bands are observed respectively. On the contrary, $0^{+}_{2}$ and $2^{+}_{2}$ show a different structure for $^{80}$Kr and $^{82}$Kr. In the latter, an axial oblate and an axial prolate $0^{+}_{2}$ and $2^{+}_{2}$ states are obtained respectively. In the former, a strong mixing between small triaxial deformed and large triaxial/prolate deformed configurations are found in the $0^{+}_{2}$ c.w.f., while the $2^{+}_{2}$ is peaked around the pure triaxial $(0.4,30^{\circ})$ shape.  
Furthermore, the axial calculations are rather consistent with the triaxial ones for $^{84-86}$Kr and for some states in $^{80-82}$Kr.\\ \indent
Finally, the structure of the neutron rich isotopes is described next. In this case, the $0^{+}_{1}$ and $2^{+}_{1}$ c.w.f. are similar between them in the $^{88-98}$Kr nuclei. Adding neutrons on top of the $N=50$ magic number produces a smooth transition from quasi-spherical shapes in $^{88}$Kr to $\gamma$-soft configurations in $^{90-92}$Kr and axial oblate shapes in $^{94-98}$Kr. Again, the evolution of the $0^{+}_{2}$ and $2^{+}_{2}$ states is more involved. In the latter, slightly deformed triaxial shapes for $^{88-94}$Kr, oblate shapes for $^{96}$Kr and prolate shapes for $^{98}$Kr are observed. The $0^{+}_{2}$ states show also different structures, namely, mostly triaxial deformation for $^{88-90}$Kr, axial oblate deformation for $^{92-96}$Kr and a prolate shape for $^{98}$Kr. It is important to point out that the c.w.f. obtained here can be directly related to the corresponding PES shown in Fig.~\ref{PES_VAP}. \\ \indent
Finally, these results are consistent with the calculations performed in the neutron rich region with the IBM method mapped to the Gogny D1M interaction~\cite{NPA_899_1_2013}.  
%%%%%%%%%%%%%%%%%%%%%%
\subsection{Systematics of the excitation energies}
%%%%%%%%%%%%%%%%%%%%%%%%
Once the shape evolution inferred from the SCCM calculations has been analyzed, the results for observables and the comparison with the available experimental data are shown. In Fig.~\ref{exc_ener}(a)-(d) the low-lying excitation energies along the isotopic chain, namely, $2^{+}_{1}$, $4^{+}_{1}$, $0^{+}_{2}$ and $2^{+}_{2}$ are plotted. A remarkable good agreement, both qualitative and quantitative, is obtained with the experimental values when the triaxial degree of freedom is taken into account, in particular for the neutron deficient $^{72-82}$Kr isotopes. The most significant differences are found around the $N=50$ magic number ($^{84-88}$Kr). Here, although the qualitative behavior of the experimental data is well reproduced, i.e., increase of the excitation energies and the maximum at $N=50$, the theoretical results overestimate the actual values. In this region, explicit quasiparticle excitations are expected to play a relevant role but they are not included in this work. \\ \indent
On the other hand, a continuous decrease of the $2^{+}_{1}$ excitation energy above $N=50$ is obtained with the triaxial calculations, revealing a smooth increase of deformation when increasing the number of neutrons. This is consistent with the shape evolution shown in the previous section. However, the experimental trend up to the last measured value ($^{96}$Kr) is flatter. In any case, neither the calculations nor the experiments support a sharp transition at $N=60$ in the Kr isotopes as the one observed in Sr and Zr isotopes (see ~\cite{NPA_899_1_2013} and references therein). \\ \indent
Concerning the possible shape coexistence expected in this region, low-lying $0^{+}_{2}$ states around 1 MeV or below in excitation energy have been measured for $^{72-78}$Kr. The triaxial calculations reproduce quite nicely these energies and they correspond to states with a strong shape mixing between prolate and oblate configurations (see Fig.~\ref{WF_ALL}). Furthermore, these calculations predict $0^{+}_{2}$ states with small excitation energies and different shapes as their corresponding ground states for $^{94,98}$Kr isotopes.\\ \indent
Finally, the results provided by the axial and triaxial calculations are compared in Fig.~\ref{exc_ener}. The axial excitation energies are much larger than the triaxial energies except for those nuclei where the triaxial c.w.f. are actually axial deformed states ($^{70-72,94-98}$Kr). Therefore, the performance of the axial calculations is significantly poorer in reproducing the experimental data.  
%%%%%%%%%%%%%%%%%%%%%%
\subsection{Systematics of the electromagnetic transition probabilities and moments}
%%%%%%%%%%%%%%%%%%%%%%%%
%%%%%%%%%%%%%%%%%%%%%%%% 
The global behavior of the electromagnetic transitions and moments along the isotopic chain is plotted in Fig.~\ref{BE2_Q21_g21_RHO0}. Only triaxial and the available experimental data are represented in this case. It is important to point out that no effective charges are used here since the valence space is very large and without a core.\\ \indent
Figure~\ref{BE2_Q21_g21_RHO0}(a) shows that the theoretical $B(E2)$ values reproduce the trend of the experimental results but they are systematically larger. Furthermore, local deviations are also found in $^{74}$Kr and, to a lesser extent, in $^{86-88}$Kr. The origin of this effect could be a slight overestimation of the deformation by the Gogny functional enhanced by the angular momentum projection. Nevertheless, the largest collectivity is observed both theoretical and experimentally around $N=40$ that indicates an erosion in this region of this harmonic oscillator shell closure. Furthermore, consistently with the behavior of the $2^{+}_{1}$ excitation energies and c.w.f., a quite smooth onset of collectivity is obtained above the $N=50$ magic number. \\ \indent
Additional information about the shape of the $2^{+}_{1}$ states is extracted from the spectroscopic quadrupole moment represented in Fig.~\ref{BE2_Q21_g21_RHO0}(b). The calculated $Q_{sp}$ values are fully consistent with the collective wave functions shown in Fig.~\ref{WF_ALL}, i.e., large $Q_{sp}(2^{+}_{1})$ positive (negative) values are obtained for the well-deformed oblate (prolate) states observed in $^{70,72,94,96,98}$Kr ($^{74,76,78}$Kr) isotopes. In addition, smaller values for those states where both the absolute deformation is small and the triaxial degree of freedom plays a role ($^{80-92}$Kr) are predicted. In such cases, the sign indicates whether the c.w.f. is more concentrated above (plus) or below (minus) $\gamma=30^{\circ}$.  
Comparing with the available data, a good agreement in the neutron deficient $^{74-78}$Kr~\cite{NPA_770_107_2006,PRC_75_054313_2007} (prolate) and stable $^{84}$Kr~\cite{PLB_546_48_2002} (slightly prolate) isotopes is found but not for the neutron rich $^{92-96}$Kr nuclei recently measured~\cite{PRL_108_062701_2012,NPA_899_1_2013}. The latter are experimentally prolate deformed while the present calculations predict an oblate character.\\ \indent
Another relevant observable that helps to analyze the interplay between collective and single-particle degrees of freedom is the gyromagnetic factor $g(2^{+}_{1})$. In the collective rotor model, this factor is approached by the simple law $g_{coll}(2^{+}_{1})=Z/A$~\cite{RING_SCHUCK}. Figure~\ref{BE2_Q21_g21_RHO0}(c) shows that, except for the nuclei close to the $N=50$ magic number, the theoretical values follow the collective model. Additionally, the experimental values are quantitatively well reproduced for $^{78-84}$Kr isotopes~\cite{PRC_64_024314_2001}, even though the nucleus $^{84}$Kr deviate significantly from the $Z/A$ trend. However, although the calculation reproduces the correct tendency, the theoretical value is largely underestimated for the semi-magic nucleus $^{86}$Kr. As it has been already mentioned above, the largest influence of quasiparticle excitations -not included here- on the nuclear structure is expected in this nucleus. \\ \indent
The last observable whose systematics along the isotopic is analyzed is the monopole electromagnetic transition strength $\rho^{2}(E0)$\cite{RMP_83_1467_2011,ADNDT_89_77_2005}. In a simplified model, where the ground and excited $0^{+}$ states are built by mixing two different intrinsic shapes, this quantity is large when both the shape mixing and the difference in the radii of the two intrinsic configurations are large~\cite{RMP_83_1467_2011,NPA_651_323_1999}. Looking at the collective wave functions shown in Fig.~\ref{WF_ALL}, these conditions are fulfilled in $^{74-82,94}$Kr. Consequently, the largest values for the $E0$ strength are obtained for such nuclei as it is plotted in Fig.~\ref{BE2_Q21_g21_RHO0}(d). However, the experimental data are only reproduced for $^{72-74}$Kr isotopes, overestimating the $\rho^{2}(E0)$ values in $^{76-82}$Kr. A plausible explanation could be that the amount of mixing provided by the present calculations in those isotopes is too large~\cite{RMP_83_1467_2011}. Nevertheless, the $E0$ strength is also quite sensitive to the precise values of the radii of the involved states and small changes in the deformation of the states can affect such a strength~\cite{RMP_83_1467_2011}. Furthermore, the explicit inclusion of fluctuations in the pairing degree of freedom can modify the final $\rho^{2}(E0)$ values~\cite{PRC_88_064311_2013}. Therefore, further studies should be performed to reproduce quantitatively the electric monopole strengths in this region.
%%%%%%%%%%%%%%%%%%%%%%
\subsection{Individual spectra}
%%%%%%%%%%%%%%%%%%%%%%
In the previous subsections, the systematics of the most relevant observables along the Krypton isotopic chain have been described. In order to give a more detailed description of the structure of each nucleus, the results of the most relevant bands obtained with the triaxial calculations are now shown and compared with the experimental values. These bands, shown in Figs.~\ref{Fig70_76}-~\ref{Fig94_98}, are built by grouping the states that are connected with the largest $B(E2)$ values. Additionally, within a given band, the structure of the collective wave functions is rather constant or evolves continuously connecting such states. Hence, the assignment of a given collective character is done by looking at the c.w.f. of the states belonging to a band. Although not all of the c.w.f. are shown here, some of them have been already discussed in Fig.~\ref{WF_ALL}.\\ \indent  
Before summarizing the results, two aspects have to be taken into account in order to provide a fair comparison between the theoretical and the experimental results. First, since neither time reversal symmetry breaking (cranking) states, explicit quasiparticle excitations nor other collective degrees of freedom such as pairing fluctuations are included, only a qualitative agreement with the experiment is expected. Proton-neutron pairing, potentially relevant around $^{72}$Kr, is not taking into account either.
Furthermore, the states that are almost pure quasiparticle excitations, like the ones expected at or near the shell closures, are out from the configuration space considered here. 
On the other hand, grouping the experimental states into bands can only be done in some nuclei, e.g., $^{72-84}$Kr, while in other isotopes such an identification is not fully clear or the data is scarce~\cite{NNDC}.\\ \indent 
Starting from the lighter nuclei, a quite similar structure of the collective bands is observed for $^{70-72}$Kr isotopes, namely, their ground state and first excited bands are built on top of an axial oblate and triaxial/prolate well deformed states respectively with a $0^{+}, 2^{+}, 4^{+}$, etc. sequence. Additionally, a second excited state triaxial band (less deformed than the other two) and a $\gamma$-band ($2^{+}, 3^{+}, 4^{+},5^{+}$, etc.) are found in both isotopes. For $^{70}$Kr, no experimental information is known while for $^{72}$Kr, the ground state band and the first excited $0^{+}$ states are measured. Compared to the theoretical results, both the $2_{1}^{+}$ and $0^{+}_{2}$ are higher in excitation energy.\\ \indent
The calculated spectra for $^{74-76-78}$Kr show again ground state and excited bands with a $\Delta I=2$ spacing and $\gamma$-bands. The overall agreement with the experimental spectra is rather good. Contrary to the $^{70-72}$Kr isotopes, the ground state bands in these nuclei are made of states with a triaxial/prolate character and a triaxial/oblate for the ones with $0^{+}_{2}$ band heads. \\ \indent
The theoretical ground state band for the nucleus $^{80}$Kr presents a triaxial character -the c.w.f. peak at $(0.3,40^{\circ})$- and the first excited band corresponds to a pseudo-$\gamma$-band with a staggering that is not present in the experimental data. Then, a $0^{+}_{2}$ state with strong shape mixing (see Fig.~\ref{WF_ALL}) connected to a triaxial band built on top of $0^{+}_{3}$ is obtained.\\ \indent
Approaching the semi magic nucleus $^{86}$Kr, the stable $^{82-84}$Kr isotopes show also in the calculations ground state and first excited bands with $\Delta I=2$ built on top of $0^{+}_{1}$ and $0^{+}_{2}$ states and a $\gamma$-band as the second excited band. For $^{82}$Kr, the ground state band and first excited are mainly formed by oblate states and prolate states, respectively, while for $^{84}$Kr is the other way around. The comparison with the experimental values is not as good as in the previous nuclei. This also happens in the closed shell nucleus $^{86}$Kr. Nevertheless, the calculations show a spherical ground state and oblate states $2^{+}_{1}, 4^{+}_{3}$ connected to it, a first excited prolate band, an yrast $4^{+}_{1}$ with mainly $K=4$ and another oblate band on top of the $0^{+}_{3}$ state. \\ \indent
For the nuclei above $N=50$ the experimental data is restricted to few states belonging basically to the ground state band. From the theoretical point of view, well-defined triaxial and oblate ground state bands are obtained for $^{88-90}$Kr and $^{94-98}$Kr isotopes respectively. For $^{92}$Kr, the triaxial ground state evolves towards oblate states when increasing the angular momentum. Additionally, $\gamma$-bands are obtained all over the nuclei above $^{86}$Kr, being the lowest in energy the one found in $^{90}$Kr. \\ \indent
Finally, shape coexistence in $^{98}$Kr isotope is predicted. In this nucleus, a clear collective spectrum is obtained, i.e., well-defined oblate ground state band -peaked at $(0.35,60^{\circ})$, triaxial/prolate first excited band -peaked at $(0.50,10^{\circ})$, and triaxial/oblate second excited band -peaked at  $(0.25,50^{\circ})$- are found. All the states belonging to the same band show practically the same c.w.f. and the excitation energies of the $0^{+}$ states are also relatively small.
For $^{94-96}$Kr isotopes, the situation is slightly different. They also have low-lying $0^{+}$ excited states and the oblate ground state bands are as well-defined as in $^{98}$Kr. However, the shape of the states of the first excited bands changes from oblate $0^{+}_{2}$ to prolate $4^{+}_{2}$ states through the $2^{+}_{2}$ states, that show triaxial/shape mixing. Since there are some fingerprints of shape coexistence in the neighboring Sr and Zr nuclei~\cite{NNDC}, more experimental data are of key importance to unveil the structural evolution of neutron rich nuclei around $N=60$.    
%%%%%%%%%%%%%%%%%%%%
\section{Summary}~\label{conclusions}
%%%%%%%%%%%%%%%%%%%%
The structure of the Krypton isotopic chain from the neutron deficient to the neutron rich nuclei has been studied with state-of-the-art SCCM methods with the Gogny D1S interaction. Beyond mean field effects have been taken into account through particle number and angular momentum projections and quadrupole shape (axial and non-axial) mixing. \\ \indent
From a mean field view, the shape evolution has been analyzed through the potential energy surfaces in the triaxial plane. Additionally, the spherical single particle shells playing a role in this region are determined from a Nilsson scheme. \\ \indent
SCCM calculations reveal a different shape evolution of the ground and excited states depending on whether the triaxial degree of freedom is included. In the full triaxial results, the ground states change from axially deformed states ($^{70-72}$Kr) to: triaxial states ($^{74-82}$Kr), a slightly deformed state ($^{84}$Kr), a spherical magic nucleus ($^{86}$Kr), slightly triaxial deformed states ($^{88-92}$Kr), and, finally, oblate states ($^{94-98}$Kr). However, for $^{74-76}$Kr the axial calculations produce oblate ground states, contrary to what is expected from the experiments~\cite{PRC_75_054313_2007}. Since the same nuclear interaction is used in both cases, the triaxial degree of freedom plays a key role to reproduce the experimental data in the neutron deficient Kr isotopes. This result confirms the ones obtained both in Ref.~\cite{PRC_74_024312_2006} and Refs.~\cite{PRC_75_054313_2007,PRC_87_054305_2013,PRC_89_054306_2014} in a unified and systematic manner.\\ \indent
The comparison with the experimental values for the first excitation energies along the isotopic chain show a nice agreement when the triaxial degree of freedom is included, specially in the neutron deficient part. However, the experimental data is only qualitatively determined around the magic nucleus $^{86}$Kr. Additionally, a continuous decrease of the $2^{+}_{1}$ excitation energies is obtained instead of the flat behavior observed experimentally. Nevertheless, the sharp transition experimentally determined in $N=60$ for Sr and Zr isotopes is not observed in the present calculations of the Krypton isotopes around this number of neutrons.\\ \indent
Concerning the electromagnetic properties, a good agreement is also obtained between the theory and the experimental data. However, some problems have been also found such as: 1) an overestimation of the actual $B(E2)$ values; 2) the results for the $Q_{sp}(2^{+}_{1})$ in $^{92-96}$Kr are in contradiction with the experimental data; and 3) an overestimation of the $E0$ strength in some nuclei. \\ \indent
Finally, collective bands of different nature (axial, spherical and triaxial deformed, with more or less shape mixing, $\gamma$-bands, etc.) have been found in the individual spectra. Shape coexistence is well reproduced in the neutron deficient isotopes, and, in addition, is predicted to appear in the nucleus $^{98}$Kr, and, to a lesser extent, in $^{94-96}$Kr isotopes.\\ \indent
As an outlook for a future work, the present results are expected to be improved by adding extra degrees of freedom in the intrinsic HFB-like basis, namely, time reversal symmetry breaking (cranking) states, parity breaking states, explicit quasiparticle excitations, proton-neutron pairing and/or other collective degrees of freedom such as pairing fluctuations. A new functional whose parameters will be adjusted by using a BMF method is also desirable. However, all of these improvements require major developments of the present SCCM method and are beyond the scope of this work.
%%%%%%%%%%%%%%%%%%%%%%%%%%%%%%%%%%%%%%%%%%%%%%%%%%%%%%%%%%%%%
\begin{acknowledgments} 
This work was supported by the Ministerio de Econom\'ia y Competitividad-Programa Ram\'on y Cajal 2012 number 11420. 
\end{acknowledgments}

\end{document}